\documentclass[preprint,11pt]{elsarticle}
\usepackage{stmaryrd,amsmath,amssymb}
\usepackage{tabularx,color}
\input psfig.sty
\usepackage[top=2.54cm,bottom=2.54cm,left=3.17cm,right=3.17cm ]{geometry}

\newcommand{\be}{\begin{equation}}
\newcommand\bj{\textbf{j}}
\newcommand\bn{\textbf{n}}
\newcommand\bm{\textbf{m}}

\newcommand\bR{\mathbb{R}}

\newcommand{\ee}{\end{equation}}
\newcommand{\ba}{\begin{array}}
\newcommand{\ea}{\end{array}}
\newcommand{\bea}{\begin{eqnarray}}
\newcommand{\eea}{\end{eqnarray}}
\newcommand{\beas}{\begin{eqnarray*}}
\newcommand{\eeas}{\end{eqnarray*}}
\newcommand{\bx}{{\bf x}}

\newcommand{\bk}{{\bf k}}
\newtheorem{exmp}{Example}
\newtheorem{remark}{Remark}[section]

\begin{document}

\begin{frontmatter}

\title{ A novel nonlocal potential solver based on nonuniform FFT for efficient simulation of the Davey-Stewartson equations} 

\author[uniMat,wpi] {Norbert J. Mauser}
\ead{norbert.mauser@univie.ac.at}
\author[uniMat,wpi,ati]{Hans Peter Stimming } 
\ead{ hans.peter.stimming@univie.ac.at}
\address[uniMat]{Faculty of Mathematics, University of Vienna, Oskar-Morgenstern-Platz 1, 1090 Vienna, Austria }
\address[wpi]{Wolfgang Pauli Institute,
University of Vienna, Oskar-Morgenstern-Platz 1, 1090 Vienna, Austria }
\address[irmar]{Universit\'e de Rennes 1, IRMAR, Campus de Beaulieu, 35042 Rennes Cedex, France}
\address[ati]{ATI, Vienna University of Technology,
Stadionallee 2, 1020 Vienna, Austria }

\author[wpi,irmar]{Yong Zhang\corref{3}}
\ead{yong.zhang@univ-rennes1.fr}

\cortext[3]{Corresponding author.}

%



\begin{abstract}
We propose an efficient and accurate solver for the nonlocal potential  in the Davey-Stewartson equations
using nonuniform FFT (NUFFT).  A discontinuity in the Fourier transform of the  nonlocal potential causes ``accuracy locking" if the potential is solved by standard FFT with periodic boundary conditions on a truncated domain. 
Using the fact that the discontinuity disappears in polar coordinates, we reformulate the potential  integral and split it into high and low frequency parts. The high frequency part can be approximated by the standard FFT method, while the low frequency part is evaluated with a high order Gauss quadrature  accelerated by nonuniform  FFT. The NUFFT solver has $O(N\log N)$ complexity, where $N$ is the total number of discretization points, and achieves higher accuracy than standard FFT solver, which makes its use in simulations very attractive. Extensive numerical results  show the efficiency and accuracy of the proposed new method.
\end{abstract}

\begin{keyword}
nonlocal potential solver, nonuniform FFT, Davey-Stewartson equations
\end{keyword}
\end{frontmatter}


\section{Introduction}
The  Davey-Stewartson (DS) equations arise as a model for free surface fluid 
waves subject to the effects of both  gravity and capillarity \cite{AbSe,CapGrav}.
A key feature of the DS equations is the nonlocal self-coupling potential term, which requires 
a suitable numerical effort. We propose in this paper a new numerical method for this nonlocal potential and  show considerable improvement gained in accuracy and efficiency over standard approaches.
  
The DS equation reads,  in dimensionless form, as follows:
\bea\label{DSII_exact} 
&&i\; \partial_t u =- \partial_{xx} u +\alpha \;\partial_{yy} u -2\; \beta\; |u|^2 u - 4\;\beta\; \Phi \;u, \quad t\ge 0, \\
\label{DSII_exact_pot} &&-\Delta \Phi = \;\partial_{xx}(|u|^2), \quad  t\ge 0,\\
&&\label{DS_initial}u(x,y,t=0) = u_0(x,y), \quad (x,y)\in {\mathbb R}^2,
\eea
where $\alpha$ can take the values $\alpha= 1$ (Hyperbolic-Elliptic DS) or $\alpha= -1$ (Elliptic-Elliptic DS), and 
the constant $\beta$ can be either positive,
representing focusing interaction, or negative, for defocusing interaction. 
 We restrict ourselves to the elliptic case of the potential coupling equation (\ref{DSII_exact_pot}),
and remark that in general, DS also allows for a hyperbolic coupling equation in place of
(\ref{DSII_exact_pot}), a case which we are not going to investigate.
In the Hyperbolic-Elliptic case ($\alpha = 1$), the equation is 
completely integrable and  called ``DSII" in the literature \cite{TheoDS}. 
This equation has recently been of interest in the modelling of rogue waves \cite{OhY}.

The nonlocal Poisson-type potential $\Phi$ in \eqref{DSII_exact_pot}, generated by the density $\rho := |u|^{2}$, can be written equivalently by Fourier transform as
\bea\label{Def_Fourier}
\Phi(\bx,t) = \int_{\mathbb R^2} -\frac{k_x^2}{k_x^2+k_y^2}\;\widehat{\rho}(\bk,t) e^{2\pi i \;\bk \cdot \bx} {\rm d} \bk,\quad\quad  \bx =(x,y) \in {\mathbb R^{2}},
\eea
where the Fourier transform is defined as follows
\begin{equation}\label{FourTransDef}
\widehat{\rho}(\bk,t) = \int_{\mathbb R^2}\rho(\bx,t)\; e^{-2\pi i \;\bk \cdot \bx} {\rm d} \bx, \quad \bk \in \mathbb R^2.
\end{equation}

In order to evaluate the nonlocal potential $\Phi$, the usual approach is to truncate the whole space problem 
onto a bounded domain and to impose some appropriate boundary conditions. 
For example, the problem can be truncated on a rectangle with periodic boundary conditions, and then a 
standard Fast Fourier Transform (FFT) method can be applied in a straightforward way \cite{BeMaSt04,KleinRoidot}. 
For the simulation of the whole coupled system, also the ``wave function'' $u$, the other unknown of DS, has to be
calculated on a bounded domain,   and the standard method is also  truncation. 
However, the domain of computation for the wave function is not necessarily the same as the one chosen for the computation of
the nonlocal potential $\Phi$. 
One might expect an accurate approximation for those $\Phi$ and $u$  that decay sufficiently fast. 
However, the discontinuity of the Fourier multiplier in \eqref{Def_Fourier}, i.e. $k_x^2/(k_x^2+k_y^2)$, at $\mathbf{k}=\mathbf{0}$, is poorly resolved if the integral is discretized directly by trapezoidal quadrature no matter how small the spatial step is, and thus causes  accuracy  locking 
similar to the one found in \cite{Bao_CPC,SPMCompare}. 
Since this discontinuity happens near low frequencies, a fast decay of the Fourier transform at far field is not necessarily leading to any smaller error.

The DS equations have been studied extensively and various aspects of numerical methods have been studied, for an incomplete list, we refer to \cite{BeMaSt04,BlowupSaut,KleinRoidot,KleinRoidotMuite}. 
A systematic numerical study of the DS equations on a torus in the semi-classical regime was done in \cite{KleinRoidot}. As pointed out above, the standard FFT applied on a bounded domain may suffer from the discontinuity and have accuracy loss in the term $\Phi$. Starting from the Fourier definition \eqref{Def_Fourier}, we will concentrate on accurate and efficient evaluation of the nonlocal potential in which the discontinuity can be resolved accurately.

We remark that also for the wave function $u$, 
the truncation to a bounded domain, together with 
periodic boundary conditions, may cause non negligible errors, for example, in the case where $u$ does not decay fast enough. 
Removing or suppressing these errors is an interesting problem on its own.
The most accurate method for this task is the Transparent Boundary Conditions (TBC). 
There are successful recent works using this method for Nonlinear Schr\"odinger equations
\cite{AntoineArnoldBesse,AntoineBesseKlein,MaScSt},
however the adaptation of these methods to cases with non-local interactions
like Davey-Stewartson is essentially  an open problem. Further work in this direction surely 
would be very important.
We mention furthermore that one could also use more flexible, but less accurate methods for
reducing domain truncation errors, like e.g. the Absorbing potential method \cite{XuHanWu}.

By adopting polar coordinates in Fourier space, the discontinuity in  \eqref{Def_Fourier} disappears automatically due to the Jacobian of the coordinate change, i.e. from Cartesian to polar coordinates, so the use of a simple high order quadrature will be possible. Here we aim to design an accurate and efficient numerical scheme which can bypass the discontinuity.  Recently, Jiang et al \cite{JLB14,BJTZ15} proposed an accurate algorithm to efficiently evaluate a series of nonlocal convolution-type potentials whose convolution kernels (and their  Fourier transforms) have singularities both in physical and Fourier space, such as Coulomb and dipolar potential in 2D and 3D, with the help of nonuniform FFT (NUFFT). 
Fortunately, the nonlocal potential $\Phi$, given by \eqref{Def_Fourier}, can be approximated in a similar way. In this paper, we adapt the nonuniform FFT (NUFFT) solver to \eqref{Def_Fourier} and combine our novel NUFFT solver with the widely used {\sl time splitting  method }
to study the dynamics of DS.

The paper is organized as follows. In Section 2, we present a detailed discussion  of the numerical algorithm together with a brief review of nonuniform FFT fast algorithms.  In Section 3, extensive numerical examples and results are shown to confirm the accuracy and efficiency of our novel NUFFT solver. Concluding remarks are drawn in Section 4.

\section{Resolving discontinuities in nonlocal potential by NUFFT algorithm}
In this section, we discuss the reasons for a new solver for the nonlocal term 
\eqref{Def_Fourier}, 
present the basic steps of the NUFFT solver, followed by a brief review of the NUFFT fast algorithm 
introduced in \cite{nufft2,nufft6} and give a  brief introduction of time splitting method for simulating the dynamics.

\subsection{Evaluation of the nonlocal potential with resolved discontinuity}\label{sec:evalPot}
As discussed above, the Fourier multiplier in the nonlocal coupling potential
\eqref{Def_Fourier} has a discontinuity at $\mathbf{k}=\mathbf{0}$, thus resolution by
a uniform quadrature does not achieve good accuracy. Moreover,
the discontinuity of the Fourier multiplier causes discontinuity in $\widehat{\Phi}$.
To be specific, from \eqref{Def_Fourier}, we have
\bea\label{Disct_Fourier}
\widehat{\Phi}(k_x,k_y) = -\frac{k_x^2}{(k_x^2+k_y^2) }\; \widehat{\rho}(k_x,k_y).
\eea
Then we have $ \widehat{\Phi}(k_x,0) =  -\widehat{\rho}(k_x,0)$ for $k_x\ne 0$, whereas 
$\widehat{\Phi}(0,k_y) = 0$ for $k_y\ne 0$.  So we obtain, in the whole space case,
\bea
 \lim_{k_x  \to 0} \widehat{\Phi}(k_x,0) =-\widehat{\rho}(\textbf{0})\quad
 \text{ and } \quad \lim_{k_y  \to 0} \widehat{\Phi}(0,k_y) = 0.
\eea
Noticing that $\widehat{\rho}(\textbf{0}) =\int_{\mathbb R^2} \rho(\bx) {\rm d}\bx \ne 0$, 
the discontinuity in $\widehat{\Phi}$ at $\bk = {\bf 0}$ is clear. The common strategy of forcing $\widehat{\Phi}(\textbf{0})=0$ in the standard FFT solver for the bounded domain problem 
results in a bad resolution of the 
discontinuity,  thus an ``accuracy locking" phenomenon is observed  \cite{Bao_CPC,SPMCompare}. 

Following a similar procedure introduced for the evaluation of nonlocal potentials in \cite{BJTZ15,JLB14}, 
we first split (\ref{Def_Fourier}) into two parts as follows:
\bea\label{PotSplitWhoSpace}
\Phi(\bx) &=& \int_{\mathbb R^2}  -\frac{k_x^2}{k_x^2+k_y^2} \widehat{\rho}(\bk) e^{2\pi i \;\bk \cdot \bx} {\rm d} \bk \\
&=&\label{SubInt1}\int_{\mathbb R^2 } \frac{-k_x^2}{k_x^2+k_y^2} \widehat{\rho}(\bk) \;(1-p(\bk)) \;e^{2\pi i \;\bk \cdot \bx} {\rm d} \bk  \\[0.05em]
&&+\label{SubInt2} \int_{\mathbb R^2 }  \frac{-k_x^2}{k_x^2+k_y^2} \widehat{\rho}(\bk) \;p(\bk) \;e^{2\pi i \;\bk \cdot \bx} {\rm d} \bk  \\[0.25em]
&:=& I_{1}(\bx) + I_{2}(\bx), \nonumber
\eea
where  $p(\bk)$ is a partition function used to separate high and low frequency parts of the
integrand.
The function $p(\bk)$ is chosen such that it satisfies two properties. First,  it has to be $C^{\infty}(\mathbb R^2)$ and to decay fast enough at far field. Second, $(1-p(\bk)) /|\bk|^2$ has to be continuous at $\bk =\textbf{0}$.  The choice of $p(\bk)$ is of course not unique,
for more details we refer to \cite{JLB14} . In this case, we choose $p(\bk) = e^{-|\bk|^2/\sigma^{2}}$ where $\sigma$ is a tunable positive number to be specified later.  Taking this $p(\bk)$, the integrand of $I_1$ is continuous in Cartesian coordinates, and the integrand of $I_2$ is continuous in polar coordinates, therefore  the discontinuity difficulty is resolved.

Since the integrand of $I_1$ is continuous and equals to zero at the origin, we can easily apply a standard FFT method.  The commonly-used technique of  setting $\widehat{\rho}(\textbf 0 ) = 0$ does not have any discontinuity difficulty in the Fourier space. To approximate $I_{1}(\bx)$ within some prescribed 
accuracy $\varepsilon$,  we can truncate the whole Fourier space into a  square, i.e. $B_f:= [-F,F]^2$ with $F =O(\varepsilon^{-1/(n-2)})$.  As is known from standard Fourier analysis, $\widehat{\rho}(\bk) = O(|\bk|^{-n})$ as $|\bk|\rightarrow \infty$ if $\rho\in C^n(\mathbb R^2)$ for positive integer $n$. 
In computation, the density is truncated on a bounded square, i.e. $B_p:=[-P,P]^{2}$, and is discretized on a uniform mesh grid $\bx_{\bn} =(x_{n_1},y_{n_2})$ with $x_{n_1} = -P+{n_1}\, h_x, n_1 =0,\ldots,N_x,  y_{n_2} = -P+ n_2\, h_y, n_2 =0,\ldots,N_y$ and $h_x = \frac{2P}{N_x}, h_y = \frac{2P}{N_y}$.
The Fourier transform $\widehat{\rho}(\bk)$ is also discretized on a uniform Fourier grid, i.e. $\bk_{\bm}=(k_{m_1},k_{m_2})$ with $k_{m_1} =\pi m_1/P,k_{m_2} =\pi m_2/P, m_1 = -N_x/2,\ldots, N_x/2+1, m_2 = -N_y/2,\ldots, N_y/2+1$.
The Fourier transforms $\widehat{\rho}(\bk_{\bm})$ are approximated by applying the trapezoidal rule to \eqref{FourTransDef} on $B_p$, and the summation is accelerated by Fast Fourier transform (FFT) within $O(N\log N) $ arithmetic operations (where $N$ is the total number of grid points). 
In implementation, we  extend the density, which is originally given on $B_{p}$, to a larger square 
$\widetilde{B_p} = [-\widetilde{P},\widetilde{P}]^{2}$ with  $\widetilde{P} =  \kappa P,  \kappa\ge 2$ by zero-padding, so as to achieve better accuracy in $I_1$ by decreasing the frequency step, i.e. $|\Delta \bk| = O(P^{-1})$.

\vspace{0.2cm}

For $I_2$, the discontinuity of the integrand is removed by the transformation to polar coordinates :
\bea
 I_2(\bx_\bn)&=&-\int_0^\infty k \;{\rm d} k  \int_0^{2\pi}  \cos^2(\theta)\widehat{\rho}(\bk) \;p(\bk) \;e^{2\pi i \;\bk \cdot \bx_\bn} \;\;{\rm d}\theta, \quad 
\bx_\bn \in B_{p}.
 \eea
Similarly to the calculation of $I_1$, $I_2$ is approximated on the bounded disk 
$D_{c}=\{|\bk|\le c,c>0\}$, where $c$ is a multiple of the parameter 
$\sigma$ in the auxiliary function $p(\bk)$.
We apply a high order Gauss-Jacobi quadrature in the radial direction with nodes and weights $
\{(r_{j_1},\omega_{j_1}^r)\}_{j_1=1}^{N_r}$, and the trapezoidal rule in the azimuthal direction with nodes and weights
$\{(\theta_{j_2},\omega_{j_2}^\theta)\}_{j_2=1}^{N_\theta}$.
Define  $\bj = (j_{1},j_{2}), \,\omega_{\bf j} = \omega_{j_1}^r \omega_{j_2}^{\theta}$ and $\bk_{\bj}=r_{j_1}(\cos\theta_{j_2},\sin\theta_{j_2})$, we have 
 \bea \label{I2PolarDisc}
 I_2(\bx_\bn) \approx - \sum_{j_1=1}^{N_r} \sum_{j_2=1}^{N_\theta}  \omega_{\bf j}
 (\cos\theta_{j_2})^{2}\;\widehat{\rho}(\bk_{\bj} )\;p(\bk_{\bj})e^{2\pi i \;\bk_{\bj} \cdot \bx_\bn}, \quad \bx_\bn\in B_p.
\eea

Note that $\bk_{\bj}$ are non-uniform grid points, therefore standard FFT can not be applied to accelerate either the computation of $\widehat{\rho}(\bk_{\bj} )$ or $I_2(\bx_\bn)$.
Direct computation of $\widehat{\rho}(\bk_{\bj} )$ or \eqref{I2PolarDisc} requires $O(N N_r N_\theta)$ operations, which is a much higher cost than a uniform FFT method requires.
Fortunately, the nonuniform FFT ``fast algorithm'' \cite{nufft2,nufft6} reduces the computational cost to $O(N\log N)$, where $N$ is the total number of grid points in both physical and Fourier space, within any prescribed approximation accuracy. The NUFFT algorithm has been developed  a lot since Dutt and Rokhlin constructed the first NUFFT algorithm with full control of precision in \cite{nufft2}.  Here we apply an optimized NUFFT proposed by Greengard and Lee in \cite{nufft6}, which utilizes Gaussians as 
interpolation or 'gridding' schemes to accelerate the algorithm by a significant factor, particularly in 2D and 3D. 
We shall not elaborate on this topic in detail, instead we refer the readers to \cite{nufft6,JLB14} for a detailed discussion. In the  section below, we shall give a really brief review of NUFFT.
 
It still remains to  decide how to choose the cutoff parameter $\sigma$ for the separation of high and low frequency regime. On one hand, one would need fewer points in the irregular, low-frequency regime $I_2$ if $\sigma$ is small, which helps reduce the computational cost of $I_2$. 
On the other hand, in this case the integrand of $I_1$ will be sharply peaked, so to achieve high accuracy a small grid step in the 
uniform grid in $I_1$ would be necessary, which would increase cost. 
Following a dimensional analysis argument, 
the parameter $\sigma$ is chosen to optimize the total computation cost as follows:
\[ \sigma = O(\min(\Delta k_x, \Delta k_y)) =  O(P^{-1}). \]
Complete details of the optimal choice are discussed and presented in \cite{JLB14}.

For the low-frequency part, we choose an irregular (polar)
$60\times 60$ grid in the Fourier domain, independently from the grid spacing in real space. This choice guarantees 12 digits of accuracy (by NUFFT) in the calculation $I_2$, a level of accuracy which is high enough (i.e. higher than error terms incurred from other parts of the algorithm) in the whole range of test cases which we will consider. The choice of the non-uniform grid step and the related accuracy were studied in \cite{JLB14}, and our choice follows recommendations given there. 

\subsection{Brief introduction of NUFFT}
As is well known, standard FFT accelerates the following summations 
\bea
&&F(k)=\sum_{j=0}^{N-1} f(j) e^{-2\pi i k j/N},\;\;k=0,\cdots,N-1,\\
&& f(j)=\frac{1}{N}\sum_{k=0}^{N-1}F(k) e^{2\pi i kj/N },\;j=0,\cdots,N-1, 
\eea
down to $O(N\log N )$ operations on uniform mesh grids in both physical domain ($\bx$) and frequency domain $\bk$.
For values defined on nonuniform mesh grids, either in physical or frequency or even in both domains, standard FFT can not be applied
due to the loss of the algebraic structure in the discrete transform matrix. NUFFT is then proposed to remove this restriction, while maintaining a computational complexity of $O(N\log N)$, where $N$ denotes the total number of points in both the physical and Fourier domains. There are three kinds of NUFFTs, namely, type 1 (from nonuniform grids to uniform grids in dual domain ), type 2 (uniform to nonuniform)  and type 3 (nonuniform to nonuniform). 

To be more specific, the type 1 NUFFT evaluates sums of the form
\begin{equation}\label{2.1}
f(\bx_j)=\sum_{n=0}^{N-1}F_n e^{i\bk_n\cdot \bx_j},
\end{equation}
for ``targets" $\bx_j$ on a uniform grid in $\bR$, given $F_n$ at nonuniform $\bk_n$ in the dual space. The type 2 NUFFT evaluates
sums of the form
\begin{equation}\label{2.2}
F(\bk_n)=\sum_{j=1}^{N} f(\bx_j) e^{-i\bk_n \cdot \bx_j},
\end{equation}
where the ``targets" $\bk_n$ are nonuniform points in $\bR$, and $f(\bx_j)$ are given on uniform grid points in the dual space. The type-3 NUFFT evaluates the sum of values defined on nonuniform grids at nonuniform grids in dual space and will not be listed here. Note that NUFFTs can be extended to higher dimensions. 

The NUFFT is essentially an approximation algorithm, which means the summations are accelerated within a controlled accuracy, and in this point it  differs from standard FFT, where efficiency is achieved by exploring the algebraic structure of the transform matrix without accuracy loss. 
The discussion of details regarding the relation of accuracy control to the algorithm parameters of NUFFT is beyond the scope of this work, and we refer to the work of Greengard et al  \cite{nufft6}.

\vspace{0.2cm}

\begin{remark}
The above algorithm can actually be applied to various nonlocal potentials once their Fourier transforms are given
analytically or can be evaluated numerically. This includes e.g. those which exist in Kadomtsev-Petviashvili equations \cite{KleinRoidot}. 
For an application to various nonlocal potentials used in BEC simulations, we refer to \cite{BJTZ15,BTZ16}. 

\end{remark}

\vspace{0.2cm}
\subsection{Time splitting method}  For computing the dynamics, we
adapt the {\sl time-splitting spectral method} (TSSP) which has been widely and successfully
used for the nonlinear Schr\"{o}dinger equation (NLSE) and DS system \cite{BeMaSt04} with many applications \cite{BaoJM,BaoJM2,BMS,SPMCompare}. We shall briefly introduce the {\sl time splitting method} for DS equations to make the following sections reasonably self-contained.
 
To advance the evolution from $t=t_n$ to $t=t_{n+1}$, we first truncate the problem (\ref{DSII_exact}) on a square $B_p$ with periodic boundary conditions for the wave function $u$ and then solve it in two steps. First we solve the linear equation, 
\bea\label{1st_tssp}
 &&i\,\partial_t v(\bx,t)=(-\partial_{xx} + \alpha\;\partial_{yy} ) v(\bx,t),\qquad \bx\in B_p,
\qquad t_n\leq t\leq
  t_{n+1} \\
  &&v(\bx,t_{n}) = u^{n}(\bx),\label{1st_tssp_more}
\eea
 for a time step $\tau$, 
  and then we solve the nonlinear equation
$t_n\leq t\leq t_{n+1}$
\bea\label{2nd_tssp}
&&i\,\partial_t w(\bx,t)=\left(-2\; \beta\; |w|^2  - 4\;\beta\; \Phi \right) w(\bx,t), \qquad \bx\in B_p,\\
&& w(\bx,t_{n}) = v(\bx,t_{n+1}),\label{2nd_tssp_more}
\eea
for the same step $\tau$, and then set $u^{n+1}(\bx)=w(\bx,t_{n+1})$.  The wave function is discretized by the Fourier spectral method, and equation (\ref{1st_tssp})-(\ref{1st_tssp_more}) can 
be integrated exactly in Fourier space.
Notice that density $\rho$ is left unchanged in (\ref{2nd_tssp}), therefore the nonlinear PDE is reduced to ODEs, pointwise in $\bx$,  and can be integrated analytically. The nonlocal potential $\Phi(\bx,t) \equiv \Phi(\bx,t_n)$ can be evaluated by the NUFFT scheme proposed in Section \ref{sec:evalPot}. 
The above time splitting scheme is the known Lie-Trotter splitting scheme. Higher order splitting methods are obtained by suitable compositions of \eqref{1st_tssp}-\eqref{1st_tssp_more} and \eqref{2nd_tssp}-\eqref{2nd_tssp_more}. 

Denote the solution to linear equation as $v^\ast := e^{\mathcal A\,\tau} u^n$ and the solution to the nonlinear equation as $u^{n+1} := e^{\mathcal B \tau} v^\ast$, where $\mathcal A$ and $\mathcal B$ are the corresponding operators. In this article, we use a fourth order time splitting method, also named Yoshida splitting method, \cite{Yoshida,KleinRoidot} to make the temporal errors negligible compared to the spatial errors, therefore we shall focus on the spatial discretization hereafter.  To be more precise, the Yoshida splitting reads as follows
\bea
u^{n+1} = e^{\mathcal A \,a_1\tau}  e^{\mathcal B\, b_1\tau} e^{\mathcal A \,a_2\tau}e^{\mathcal B\, b_2\tau}  
e^{\mathcal A \,a_3\tau} e^{\mathcal B \,b_3 \tau} e^{\mathcal A \,a_4\tau}\, u^n,
\eea
with the coefficients 
$$b_3 = \frac{1}{\small{2-\sqrt[3]{2}}}=b_1,\,b_2 = -\sqrt[3]{2} \,b_3, \;\;a_4 = \frac{1}{2} b_3 = a_1,\, a_3 = \frac{1}{2} (b_3+ b_2) = a_2.$$

\section{Numerical results}
In this section, we shall present numerical results  concerning the accuracy and efficiency of our novel NUFFT solver.
We apply the pseudo-spectral Fourier method on the rectangle $[-L_x,L_x]\times [-L_y,L_y]$ with periodic 
boundary conditions. We denote the mesh step by $h_x = 2L_x/N_x, h_y = 2L_y/N_y$ and the mesh grid by $(x_j,y_k) := (-L_x + j h_x, -L_y + k h_y),\;
j = 0,1,\cdots, N_x, \;k =0,1,\cdots, N_y$ with $N_x,N_y$ being positive even integers. For the sake of simplicity, we choose 
$L_x = L_y, \; h_x =h_y$ and denote them by $L$ and $h$ respectively. 


\subsection{Comparison for nonlocal potential evaluation} 

In this subsection, we shall present accuracy and efficiency results of both 
standard FFT and NUFFT solver for two different examples of calculation of $\Phi$ for a given $\rho$. 
In the  tables of this subsection, we use the relative discrete $l^{\infty}$ norm 
as error measure, i.e. $\|\Phi_{L,h} -\Phi\|_{l^{\infty}}/ \|\Phi\|_{l^\infty}$.

\begin{exmp}\label{Gaussian_examp} Gaussian density.\end{exmp} 

We choose $\rho(x,y) = \pi \;e^{-\pi^2 (x^2+y^2)}$, and the potential is given explicitly, in polar coordinates, as
\bea
\Phi(r,\theta)=  -\left(\frac{\pi}{2}\;e^{-\pi ^2 r^2} +\cos(2 \theta )\;  e^{-\pi ^2 r^2} (2\pi r^2)^{-1}(1+\pi ^2 r^2-e^{\pi ^2 r^2})\right).
\eea

First, we present the accuracy results of FFT and NUFFT. Table \eqref{tab:stdFFT_Gaussian} presents the errors of the potential obtained by the standard FFT solver on $[-L,L]^2$ with different mesh sizes $h$  and  errors with the same mesh $h = 1/16$ on different domains.
Table (\ref{tab:NUFFT_Gaussian}) shows errors of the potential obtained by the NUFFT solver on $[-L,L]^2$ with different mesh size $h$.

Second, the efficiency of NUFFT is shown in terms of CPU time.
Table \eqref{tab:NUFFT_Gaussian_time} displays the errors and computational cost of the 
NUFFT solver on $[-32,32]^2$. The algorithm is implemented in Fortran, the code is compiled by 
ifort 13.1.2 using the option -O3, and  executed on 32 bit Ubuntu Linux on a 2.90GHz Intel(R) Core(TM) i7-3520M CPU with 6MB cache.

Thirdly, we compare NUFFT and FFT in terms of computation time needed to achieve a given accuracy.  
We just report one comparison result for the sake of brevity. It takes $0.12$ seconds for the FFT solver on $[-32,32]^2$ with $h= 1/16$ and the achieved accuracy is $1.49$E$-04$. For the NUFFT solver on $[-8,8]^2$ with $h = 1/4$, a time of $0.02$ seconds is needed, and the  obtained accuracy is $2.53$E$-04$, which is comparable in size to the accuracy of the previous case. So the new method clearly achieves a time benefit.

From Tables (\ref{tab:stdFFT_Gaussian})-(\ref{tab:NUFFT_Gaussian_time}), we can conclude that:
(i) The accuracy by the standard FFT solver reaches a saturation as the mesh size decreases, where the errors coming from the periodic boundary condition approximation dominates; (ii) For the standard FFT solver, the boundary condition error decreases as the domain increases, and the convergence rate with respect to volume of the domain is one; (iii) The NUFFT solver approximates the potential with spectral accuracy and the computational cost is $O(N \log(N)$) 
(with $N$ being the total number of grid points in physical domain). 
The accuracy of this solver is better than the best accuracy achieved by the standard FFT solver on a much larger domain. 
Thus the new solver is quite efficient, which shall make it an ideal choice in long time dynamics of DS.

\begin{table}[t!]
\tabcolsep 0pt \caption{Errors of the potential by standard FFT solver in Example \ref{Gaussian_examp}.}
\label{tab:stdFFT_Gaussian}
\begin{center}\vspace{-1em}
\def\temptablewidth{1\textwidth}
{\rule{\temptablewidth}{1pt}}
\begin{tabularx}{\temptablewidth}{@{\extracolsep{\fill}}p{1.35cm}rllll}
&  $h = 1$  & $h= 1/2$ & $h=1/4$ & $h  = 1/8$ & $h = 1/16$ \\[0.25em]\hline
$L = 8$ & 2.08E-01	&2.22E-02	&2.38E-03	&2.38E-03	&2.38E-03 \\
$L = 16$& 2.05E-01	&2.12E-02	&5.95E-04	&5.95E-04	&5.95E-04\\
[0.25em]\hline
	&$L = 8$  & $L = 16$ & $L = 32$ & $L = 64$ & $L = 128$ \\[0.25em]\hline
 $h \!\!=1/16\!\!$& 2.38E-03& 5.95E-04 & 1.49E-04 & 3.72E-05 &9.30E-06 \\
\end{tabularx}
{\rule{\temptablewidth}{1pt}}
\end{center}
\end{table}

\begin{table}[t!]
\tabcolsep 0pt \caption{Errors of the potential by the NUFFT solver in Example \ref{Gaussian_examp}.}
\label{tab:NUFFT_Gaussian}
\begin{center}\vspace{-1em}
\def\temptablewidth{1\textwidth}
{\rule{\temptablewidth}{1pt}}
\begin{tabularx}{\temptablewidth}{@{\extracolsep{\fill}}p{1.35cm}rllll}
&  $h = 1$  & $h= 1/2$ & $h=1/4$ & $h  = 1/8$ & $h = 1/16$ \\[0.25em]\hline
$L = 8$ & 2.04E-01	&2.08E-02&	2.53E-04	&1.56E-10&	2.69E-15\\
$L = 16$&2.04E-01	&2.08E-02&	2.53E-04	&1.56E-10&	2.83E-15\\
\end{tabularx}
{\rule{\temptablewidth}{1pt}}
\end{center}
\end{table}

\begin{table}[t!]
\tabcolsep 0pt 
\begin{center}
\caption{Errors and computational time of Example \ref{Gaussian_examp} on $[-32,32]^2$.
 $T_{HF}$: time for high frequency part $I_1$ in (\ref{PotSplitWhoSpace}), 
$T_{NUFFT}$: time for NUFFT part $I_2$ in (\ref{PotSplitWhoSpace}).}
\label{tab:NUFFT_Gaussian_time}
\vspace{-1em}
\def\temptablewidth{1\textwidth}
{\rule{\temptablewidth}{1pt}}
\begin{tabularx}{\temptablewidth}{@{\extracolsep{\fill}}p{1.35cm}rlll}
 & $T_{HF}$ & $T_{NUFFT}$ & $T_{Total}$ &$E$ \\
$h  = 1/2$&2.00E-02&1.00E-02&3.00E-02&2.08E-02\\
$h  = 1/4$&9.00E-02&4.00E-02&1.30E-01&2.53E-04\\
$h  = 1/8$&3.70E-01&1.50E-01&5.20E-01&1.56E-10\\
$h\!\!=1/16\!\!$&1.66&5.80E-01&2.24&2.69E-15\\
\end{tabularx}
{\rule{\temptablewidth}{1pt}}
\end{center}
\end{table}

\vspace{0.5cm}

\begin{exmp}\label{Arkadiev_examp}Arkadiev's  exact solution with geometric decay. \end{exmp}

In the exact solution given by Arkadiev \cite{Arkadiev},
the density and the corresponding potential are given explicitly as follows
\bea
\rho(\bx) = \frac{4}{((x+1)^2+y^2+1)^2}, \quad \Phi(\bx) = -2\;\; \dfrac{y^2-(x+1)^2+1}{((x+1)^2+y^2+1)^2}.
\eea
Note that density  and the potential both decay geometrically. It is known that  these rational decay solutions are notoriously hard to simulate numerically due to their slow decay at far field, which in turn requires large computational domains so as to achieve satisfactory results.

We present  the accuracy results by the standard FFT and NUFFT solver and the efficiency performance of NUFFT. 
Table \eqref{tab:stdFFT_Arkadiev} presents errors of the potential obtained by the standard FFT solver on $[-L,L]^2$ 
with different $h$. Table (\ref{tab:NUFFT_Arkadiev}) shows errors of the potential by the NUFFT solver on $[-L,L]^2$.
Table (\ref{tab:NUFFT_Arkadiev_time}) displays errors and computational cost of the 
NUFFT solver on $[-64,64]^2$ on the same computer as in Example \ref{Gaussian_examp}.

Also in this example, from Tabs. (\ref{tab:stdFFT_Arkadiev})-(\ref{tab:NUFFT_Arkadiev_time}) we can observe a saturation due to the truncation and the periodic boundary condition approximation for FFT solver. The NUFFT solver does not perform as well as in Example \ref{Gaussian_examp}, and the main reason should be the geometric decay of the potential and density. However, we could still achieve much better accuracy than standard FFT for equal discretization step and domain size,
i.e. several orders of magnitude for the finest resolution tested. Thus the NUFFT solver is still an ideal choice for the simulation of long time dynamics.

\begin{table}[t!]
\tabcolsep 0pt \caption{Errors of the potential by standard FFT solver in Example \ref{Arkadiev_examp}.}
\label{tab:stdFFT_Arkadiev}
\begin{center}\vspace{-1em}
\def\temptablewidth{1\textwidth}
{\rule{\temptablewidth}{1pt}}
\begin{tabularx}{\temptablewidth}{@{\extracolsep{\fill}}p{1.35cm}rllll}
&  $h = 1$  & $h= 1/2$ & $h=1/4$ & $h  = 1/8$ & $h = 1/16$ \\[0.25em]\hline
 $L = 8$ &3.40E-02&2.51E-02&2.63E-02&2.69E-02&2.73E-02\\
$L = 16$ &2.44E-02&6.11E-03&6.24E-03&6.31E-03&6.34E-03\\
$L = 32$ &2.20E-02&1.50E-03&1.51E-03&1.52E-03&1.53E-03\\
$L = 64$ &2.13E-02&4.41E-04&3.73E-04&3.74E-04&3.74E-04\\
\end{tabularx}
{\rule{\temptablewidth}{1pt}}
\end{center}
\end{table}

\begin{table}[t!]
\tabcolsep 0pt \caption{Errors of the potential by NUFFT solver 
in Example \ref{Arkadiev_examp}.}
\label{tab:NUFFT_Arkadiev}
\begin{center}\vspace{-1em}
\def\temptablewidth{1\textwidth}
{\rule{\temptablewidth}{1pt}}
\begin{tabularx}{\temptablewidth}{@{\extracolsep{\fill}}p{1.35cm}rllll}
&  $h = 1$  & $h= 1/2$ & $h=1/4$ & $h  = 1/8$ & $h = 1/16$ \\[0.25em]\hline
$L = 8$  &2.12E-02&2.51E-04&8.27E-05&8.76E-05&9.02E-05\\
$L = 16$ &2.12E-02&2.49E-04&4.28E-06&4.40E-06&4.46E-06\\
$L = 32$ &2.11E-02&2.49E-04&1.11E-06&2.47E-07&2.48E-07\\
$L = 64$ &2.11E-02&2.49E-04&1.11E-06&1.46E-08&1.47E-08\\
\end{tabularx}
{\rule{\temptablewidth}{1pt}}
\end{center}
\end{table}

\begin{table}[t!]
\tabcolsep 0pt \caption{Errors and computational time  of Example \ref{Arkadiev_examp} on $[-64,64]^2$. }
\label{tab:NUFFT_Arkadiev_time}
\begin{center}\vspace{-1em}
\def\temptablewidth{1\textwidth}
{\rule{\temptablewidth}{1pt}}
\begin{tabularx}{\temptablewidth}{@{\extracolsep{\fill}}p{1.35cm}rlll}
 & $T_{HF}$ & $T_{NUFFT}$ & $T_{Total}$ &$E$ \\
$h  = 1$&0.04&0.01&0.05&2.11E-02 \\
$h  = 1/2$&0.18&0.03&0.21&2.49E-04\\
$h  = 1/4$&0.96&0.12&1.08&1.11E-06\\
$h= 1/8$&4.90&0.55&5.45&1.46E-08\\
\end{tabularx}
{\rule{\temptablewidth}{1pt}}
\end{center}
\end{table}

\subsection{ Comparison for Hyperbolic-Elliptic DS simulation}

As shown in the last subsection, the NUFFT solver proves to be accurate and efficient for evaluation
of the nonlocal potential.
In this subsection, we compare these two solvers in terms of accuracy when applied to simulate the 
complete dynamics of the DS equations (\ref{DSII_exact})-(\ref{DS_initial}).
In this subsection, we use the relative discrete $l^2$ norm 
as error measure for all reported examples.

\begin{exmp}\label{Hyper_Ellip}Hyperbolic-Elliptic DS equation.  \end{exmp}
We first study the defocusing H-E Case with  $\alpha = 1$, $\beta=-1$.
In this case, we consider the following two initial values:
\begin{enumerate}
\item {\sl Case I (Arkadiev's solution)}. We consider again the exact solution given in \cite{Arkadiev}. 
The initial value  is 
\bea
u_0(\bx) =  \dfrac{2 e^{i 2y }}{(x+1)^2+ y^2+1}, \eea 
and for the evolution,
\bea
&&u(x,y,t)= \!\dfrac{2 e^{i (2y+ 4t)} }{(x+1)^2+(y+4t)^2\!+\!1}, \\
&& \Phi(x,y,t) \!=\!-2\dfrac{(y+4t)^2-x(x+2)}{\left((x+1)^2+(y+4t)^2+1\right)^2}.
\eea

\item {\sl Case II (Gaussian initial value)}. Consider a Gaussian initial value as
\bea u_0(x,y) = e^{-\frac{(x^2+y^2)}{4}}.\eea
We use a highly resolved reference solution, obtained  by applying a  {\sl fourth-order time splitting Fourier pseudo spectral method} (TSFP4)  \cite{Yoshida} coupled with NUFFT potential solver on $[-64,64]^2$ with a fine mesh size $h = 1/8$ and time step $\Delta t =0.001$.
\end{enumerate}

We truncate the problem onto a bounded domain, i.e. a rectangle $[-L,L]^2$, and 
 impose  periodic boundary conditions for $u$, which could be well resolved by the Fourier pseudo-spectral method. The integrator was chosen as fourth-order time splitting method \cite{Yoshida} with a small time step, i.e. $\Delta t = 0.001$, such that errors coming from the temporal discretization are negligible. The potential $\Phi$ is evaluated by FFT or NUFFT solver on the same uniform mesh grid as the wave function.
For brevity, we shall denote the TSFP4 coupled with FFT potential solver by FFT, and TSFP4 coupled with NUFFT potential solver by NUFFT.

Table (\ref{dy:heds:case1:fft}) and  (\ref{dy:heds:case1:nufft}) present  errors of the wave function and potential at time $T = 0.4$ obtained by FFT and NUFFT for case I.
Table (\ref{dy:heds:case2:fft}) and  (\ref{dy:heds:case2:nufft}) show errors of the wave function and potential at time $T = 0.4$ obtained by FFT and NUFFT for case II.

From Tabs. (\ref{dy:heds:case1:fft})-(\ref{dy:heds:case2:nufft}), we can draw the following conclusions: 
(i) To achieve a fair accuracy in the potential by FFT solver, one needs to choose a very large domain; while the NUFFT solver essentially does not depend on the size of the computational domain. Moreover it is spectrally accurate provided the wave function decays fast enough.  (ii) The accuracy for the wave function in case I does not improve too much by NUFFT. Here the error coming from the the bounded domain truncation of $u$ is dominant. However, all the simulations done with NUFFT show better results than the FFT method. To significantly increase accuracy in this case, a treatment of boundary cutoff errors for the wave function $u$ is necessary; this direction of work will be the subject of a future paper.
For case II, where the wave function decays exponentially fast, the improvement  of the NUFFT solver is obviously observed.

\vspace{0.2cm}

\begin{table}[t!]
\tabcolsep 0pt 
\caption{Errors of the wave function (upper) and potential (below) by FFT solver for Example \ref{Hyper_Ellip}: case I. }
\label{dy:heds:case1:fft}
\begin{center}\vspace{-1em}
\def\temptablewidth{1\textwidth}
{\rule{\temptablewidth}{1pt}}
\begin{tabularx}{\temptablewidth}{@{\extracolsep{\fill}}p{1.35cm}rlll}
&  $h = 1$  & $h= 1/2$ & $h=1/4$ & $h  = 1/8$  \\[0.25em]\hline
 $L = 8$   &3.30E-01&5.36E-02&4.82E-02&4.85E-02\\
 $L = 16$  &3.30E-01&3.14E-02&1.73E-02&1.73E-02\\
 $L = 32$  &3.30E-01&2.78E-02&8.25E-03&8.26E-03\\
 $L = 64$  &3.30E-01&2.68E-02&2.23E-03&2.23E-03\\
[0.25em]\hline
 $L = 8$   &2.54E-01&1.63E-01&1.63E-01&1.64E-01\\
 $L = 16$  &2.22E-01&8.24E-02&8.17E-02&8.17E-02\\
 $L = 32$  &2.14E-01&4.23E-02&4.08E-02&4.08E-02\\
 $L = 64$  &2.12E-01&2.33E-02&2.04E-02&2.04E-02\\
\end{tabularx}
{\rule{\temptablewidth}{1pt}}
\end{center}
\end{table}

\begin{table}[t!]
\tabcolsep 0pt \caption{Errors of the wave function (upper) and potential (below) by NUFFT for Example \ref{Hyper_Ellip}: case I. }
\label{dy:heds:case1:nufft}
\begin{center}\vspace{-1em}
\def\temptablewidth{1\textwidth}
{\rule{\temptablewidth}{1pt}}
\begin{tabularx}{\temptablewidth}{@{\extracolsep{\fill}}p{1.35cm}rlll}
&  $h = 1$  & $h= 1/2$ & $h=1/4$ & $h  = 1/8$  \\[0.25em]\hline
 $L = 8$   &3.35E-01&3.81E-02&2.82E-02&2.88E-02\\
 $L = 16$  &3.31E-01&3.02E-02&1.42E-02&1.42E-02\\
 $L = 32$  &3.30E-01&2.78E-02&7.88E-03&7.89E-03\\
 $L = 64$  &3.30E-01&2.68E-02&2.15E-03&2.15E-03\\
[0.25em]\hline
 $L = 8$   &2.00E-01&1.12E-02&1.91E-03&1.88E-03\\
 $L = 16$  &2.08E-01&1.13E-02&4.52E-05&1.72E-04\\
 $L = 32$  &2.10E-01&1.13E-02&5.84E-06&4.32E-06\\
 $L = 64$  &2.11E-01&1.13E-02&4.15E-06& 5.06E-07\\
\end{tabularx}
{\rule{\temptablewidth}{1pt}}
\end{center}
\end{table}

\begin{table}[t!]
\tabcolsep 0pt \caption{Errors of the wave function (upper) and potential (below) by FFT for Example \ref{Hyper_Ellip}: case II. }
\label{dy:heds:case2:fft}
\begin{center}\vspace{-1em}
\def\temptablewidth{1\textwidth}
{\rule{\temptablewidth}{1pt}}
\begin{tabularx}{\temptablewidth}{@{\extracolsep{\fill}}p{1.35cm}rlll}
&  $h = 1$  & $h= 1/2$ & $h=1/4$ & $h  = 1/8$  \\[0.25em]\hline
 $L = 8$   &2.03E-02  & 1.96E-02  & 1.96E-02  & 1.96E-02\\
 $L = 16$  &7.16E-03  & 4.91E-03  & 4.91E-03  & 4.91E-03\\
 $L = 32$  &5.36E-03  & 1.23E-03  & 1.23E-03  & 1.23E-03\\
 $L = 64$  &5.22E-03  & 3.07E-04  & 3.07E-04  & 3.07E-04\\ [0.25em]\hline
 $L = 8$   &1.91E-01 & 1.91E-01 &  1.91E-01  & 1.91E-01\\
 $L = 16$  &9.55E-02 & 9.53E-02 &  9.53E-02  & 9.53E-02\\
 $L = 32$  &4.79E-02 & 4.77E-02 &  4.77E-02  & 4.77E-02\\
 $L = 64$  &2.42E-02 & 2.38E-02 &  2.38E-02  & 2.38E-02\\
\end{tabularx}
{\rule{\temptablewidth}{1pt}}
\end{center}
\end{table}

\begin{table}[t!]
\tabcolsep 0pt \caption{Errors of the wave function (upper) and potential (below)  by NUFFT for Example \ref{Hyper_Ellip}: case II. }
\label{dy:heds:case2:nufft}
\begin{center}\vspace{-1em}
\def\temptablewidth{1\textwidth}
{\rule{\temptablewidth}{1pt}}
\begin{tabularx}{\temptablewidth}{@{\extracolsep{\fill}}p{1.35cm}rlll}
&  $h = 1$  & $h= 1/2$ & $h=1/4$ & $h  = 1/8$  \\[0.25em]\hline
 $L = 8$   &5.14E-03  & 8.03E-06  & 5.09E-07  & 4.69E-07\\
 $L = 16$  &5.21E-03  & 8.07E-06  & 3.26E-11  & 3.09E-13\\
 $L = 32$  &5.21E-03  & 8.01E-06  & 3.26E-11  & 5.02E-13\\
[0.25em]\hline
 $L = 8$   &3.34E-03 &  2.01E-05  & 1.90E-08  & 1.28E-08\\
 $L = 16$  &3.85E-03 &  2.05E-05  & 8.84E-11  & 4.22E-13\\
 $L = 32$  &3.98E-03 &  2.07E-05  & 8.91E-11  & 8.11E-14\\
\end{tabularx}
{\rule{\temptablewidth}{1pt}}
\end{center}
\end{table}

\subsection{ Comparison for Elliptic-Elliptic DS simulation}
Here we take the same example as treated in \cite{BlowupSaut,BeMaSt04}.
All the errors presented in this subsection are relative discrete $l^2$ errors.
\begin{exmp}\label{Ellip_Ellip}Elliptic-Elliptic DS equation. \end{exmp}
\beas
&&i\, \partial_t u = -(\partial_{xx} +\partial_{yy}) \;u + \chi\; |u|^{2} u + u\; \phi,\quad \bx\in {\mathbb R}^{2},\\
&&-(\partial_{xx} +\partial_{yy}) \phi = \gamma \; |u|^{2}_{xx},
\eeas
with initial value chosen as a Gaussian, i.e. $u(\bx,0) =u_0(\bx)= 4\; e^{-\frac{x^2+y^2}{4}}$. 
In this example, it is known that finite time blowup can occur. 
We choose a maximal computation time smaller than the time of blow-up
to test the accuracy performance of the NUFFT solver.  We take $\chi =-1$ (focusing interaction), $\gamma =1$ and final time $T = 0.05$, which is smaller than the  blow up critical time $T_\ast \approx 0.1311$. 
The reference solution used to measure the errors is obtained numerically by applying TSFP4 coupled with the NUFFT solver on the domain $[-64,64]^2$ with a fine mesh size $h = 1/8$ and very fine time step $\Delta t =0.0001$ such that errors coming from the temporal discretization are negligible.

Table \eqref{dy:ee:fft} and \eqref{dy:ee:nufft} present errors of the wave function and potential at time $T = 0.05$ obtained by FFT and NUFFT respectively, from which we can draw similar conclusions as in Example \ref{Hyper_Ellip}.

\begin{table}[t!]
\tabcolsep 0pt \caption{Errors of the wave function (upper) and potential (below) by standard FFT for Example \ref{Ellip_Ellip}.}
\label{dy:ee:fft}
\begin{center}\vspace{-1em}
\def\temptablewidth{1\textwidth}
{\rule{\temptablewidth}{1pt}}
\begin{tabularx}{\temptablewidth}{@{\extracolsep{\fill}}p{1.35cm}rlll}
&  $h = 1$  & $h= 1/2$ & $h=1/4$ & $h  = 1/8$  \\[0.25em]\hline
$L = 8$  &1.26E-02  & 9.82E-03 &  9.82E-03  & 9.82E-03\\
$L = 16$ &6.95E-03  & 2.46E-03 &  2.45E-03  & 2.45E-03\\
$L = 32$ &6.16E-03  & 6.17E-04 &  6.14E-04  & 6.14E-04\\
[0.25em]\hline
$L = 8$  &1.84E-01  & 1.83E-01 &  1.83E-01  & 1.83E-01\\
$L = 16$ &9.25E-02  & 9.14E-02 &  9.14E-02  & 9.14E-02\\
$L = 32$ &4.74E-02  & 4.57E-02 &  4.57E-02  & 4.57E-02\\
\end{tabularx}
{\rule{\temptablewidth}{1pt}}
\end{center}
\end{table}

\begin{table}[t!]
\tabcolsep 0pt \caption{Errors of the wave function (upper) and potential (below) by NUFFT for Example \ref{Ellip_Ellip}.}
\label{dy:ee:nufft}
\begin{center}\vspace{-1em}
\def\temptablewidth{1\textwidth}
{\rule{\temptablewidth}{1pt}}
\begin{tabularx}{\temptablewidth}{@{\extracolsep{\fill}}p{1.35cm}rlll}
&  $h = 1$  & $h= 1/2$ & $h=1/4$ & $h  = 1/8$  \\[0.25em]\hline
$L = 8$  &5.88E-03  & 6.15E-05 &  7.52E-08 &  6.05E-08 \\
$L = 16$ &5.97E-03  & 6.21E-05 &  3.17E-08 &  5.59E-13 \\
$L = 32$ &5.99E-03  & 6.22E-05 &  3.18E-08 &  7.37E-13 \\
[0.25em]\hline
$L = 8$   &1.18E-02 &  9.52E-05 &  1.06E-07 &  9.60E-11\\
$L = 16$  &1.22E-02 &  9.82E-05 &  1.08E-07 &  7.08E-13\\
$L = 32$  &1.23E-02 &  9.90E-05 &  1.08E-07 &  2.52E-13\\
\end{tabularx}
{\rule{\temptablewidth}{1pt}}
\end{center}
\end{table}

\section{Conclusion}
We propose an efficient and accurate solver for the nonlocal potential in Davey-Stewartson system
utilizing a nonuniform FFT (NUFFT) fast algorithm. 
We show in this paper that the NUFFT solver helps to achieve much better accuracy in the calculation of the nonlocal coupling term while still maintaining $O (N \log N)$ complexity, the same as uniform FFT. 
Extensive numerical results are presented to show its efficiency and accuracy. 

For a wave function with exponential decay, the new solver reaches machine precision, while for the lump solution according to Arkadiev with algebraic decay, there still is a significant gain in accuracy. For the latter case, the dominant error comes from the truncation of the wave function and the advantage of NUFFT is masked.
In this case, the NUFFT solver should be coupled to a method for removing cutoff errors of the wave function $u$, and future work in this direction is planned.
The dramatic improvement in the nonlocal potential  evaluation also benefits greatly the simulation of the DS equations. 
The gain in accuracy and efficiency makes the new solver a good alternative in simulations.
Further extensions of the NUFFT solver to other nonlocal potentials are possible.

\section*{Acknowledgments}
The authors acknowledge support from the ANR-FWF Project LODIQUAS (ANR-11-IS01-0003 and FWF grant No I830) and the Austrian Science Foundation (FWF) under grant No F41 (project VICOM), the European Research Council (ERC) under starting grant 258603 NAC, and the Austrian Ministry of Science
and Research via its grant for the WPI. Y. Zhang acknowledges support from ANR project Moonrise ANR-14-CE23-0007-01 and the Natural Science Foundation of China grants  11261065, 91430103 and 11471050.
The computational results presented have been achieved using the Vienna Scientific Cluster (VSC).


\end{document}